\documentclass[12pt]{JHEP3}
\usepackage{latexsym} 
\usepackage{epsfig,amssymb,euscript}
\usepackage{amsmath} 

\def\be{\begin{equation}} \def\ee{\end{equation}}
\def\bea{\begin{eqnarray}} \def\eea{\end{eqnarray}}
\newcommand{\nn}{\nonumber}

\newcommand{\pa}{\partial}

\title{
{\begin{center}
Soft Spectrum in Yukawa-Gauge Mediation
\end{center}}
}
\author{Federico Galli$^{a}$, Alberto Mariotti$^{b}$
~~\\
Theoretische Natuurkunde, Vrije Universiteit Brussel, \\
and International Solvay Institutes,\\ 
Pleinlaan 2, B-1050 Brussels, Belgium\\
 ~~\\
  $^a$\email{federico.galli@vub.ac.be} ~~~
$^b$\email{alberto.mariotti@vub.ac.be}
}

\abstract{We introduce a model independent parametrization
for a subclass of gauge mediated theories, which we refer to as
Yukawa-gauge mediation.
Within this formalism we study the resulting
soft masses in the visible spectrum.
We find general expressions for the gaugino
and scalar masses. 
Under generic conditions,
the gaugino mass is screened,
vanishing
at first order in the SUSY breaking scale.}
\begin{document}

\maketitle

\section{Introduction}
During the last few years there has been fervent activity in the study of the phenomenology
of gauge mediated models 
of supersymmetry breaking in the SSM 
(see \cite{Martin,Giudice} for reviews and references).
The recent 
developments  in finding calculable and
metastable vacua with dynamical supersymmetry breaking \cite{ISS}
have opened up the possibility of building models for gauge mediated scenarios which
are not only predictive and weakly coupled, but also have a dynamical UV completion.
Typically these models reduce in the infrared to effective weakly coupled models 
of pure chiral fields which are generalization of O'Raifeartaigh models,
whose phenomenology can be studied 
(see e.g. \cite{Extraordinary}).

With the incoming data of the LHC, 
it is relevant to study the signature of  specifics mechanisms of mediation of supersymmetry breaking, irrespectively
of the intricate details of the hidden sector.
This has been the approach of general gauge mediation (GGM)
\cite{GGM} and subsequent works \cite{Generici} 
(see for instance \cite{pheno} for some collider
studies on the GGM parameter space).
In particular, it is interesting to investigate the generic structure and hierarchies
in the resulting soft spectrum, in order to have smoking guns 
indicating particular classes of models
in the complicated analysis of the soft  SSM 
parameters.\footnote{In this context $R$-symmetry plays a crucial role. If unbroken, it forbids gaugino mass generation.
Moreover, it has been recently shown that $R$-symmetry can
further control the suppression of gaugino mass in general chiral models, 
leading to important no go theorems \cite{Shih}.
}

The analysis of GGM \cite{GGM} has identified the complete parameter
space that is allowed by gauge mediation.
A basic effective model, which already 
cover a relevant portion of the parameter space of  general gauge mediation,
is minimal gauge mediation.
In minimal gauge mediation a pair of vector-like chiral fields,
the messengers, are charged under the SSM gauge groups and
 couple in the superpotential to a spurion singlet superfield $X$, whose
scalar and $F$-term components take expectation values.
This simple structure is  
realized, typically with some extra superfields and interactions, 
in many effective models of metastable dynamical 
supersymmetry breaking (see \cite{SeibergIntriligator} and references therein).

In this paper we investigate a natural generalization of this scenario which
we argue can often emerge as the effective theory in models  with dynamical supersymmetry breaking.
It is indeed natural to generalize minimal gauge mediation by replacing the spurion singlet with 
a dynamical field $X$ and coupling the latter, via superpotential interaction only, to some hidden sector chiral operator. 
In this setup, we will assume that the field $X$ does not get any tree level scalar or $F$-term vacuum expectation value. The  superfield $X$ senses the supersymmetry breaking  at loop level via the superpotential interaction. In the limit in which this interaction is switched off the visible sector has a complete supersymmetric spectrum.
 These assumptions make the setup substantially different from minimal gauge mediation and heavily affect  the consequent phenomenology.

Concrete models realizing similar scenarios have been, for instance, considered in \cite{Kang:2010ye,Trieste}.
The purpose of this paper is to provide a general analysis of the resulting 
soft spectrum in the SSM for  this class of models, 
here referred to as Yukawa-gauge mediation.

Note that our scenario is included in GGM \cite{GGM} 
and also in the general messenger gauge mediation of 
\cite{GMGM}. However, in our working assumption of vanishing $F$-term for $X$, 
the contribution to the gaugino mass computed in \cite{GMGM} is not present. 
Hence
our analysis deals with the next order correction to the result of \cite{GMGM}.

Another more technical motivation to study Yukawa-gauge mediation is that it may provide
 a class of models where the ratio between the
gaugino and scalar masses is of one loop in a weak coupling expansion.
Many  models have gauginos which are lighter than the scalars because of
$F$-term suppression, e.g. \cite{terzordine}. On the contrary, a one loop order hierarchy cannot
be easily realized in general. 
Semi-direct gauge mediation \cite{SemiDirect} would be a natural candidate, but it suffers from gaugino screening \cite{Hamed,Argurio,Cohen:2011aa}.

Our computation shows that also in Yukawa-gauge mediation the gaugino mass is screened.
This agrees with and generalizes the results of  \cite{Hamed,Giudice2},  obtained using wave function renormalization techniques.  
Interestingly, we find that the gaugino mass screening is 
realized as a suppression in powers of the SUSY breaking scale and not in loop factors.
Therefore this class of models gives a further peculiar realization of the gaugino screening phenomenon.

The organization of the paper is the following. In the next section we introduce the model 
and we set our parametrisation for the hidden sector. In section \ref{gaugino}
we work out the expression of the visible gaugino mass, 
compute the two loops integrals involved  and discuss the resulting suppression. 
Then we add a 
superpotential term 
that breaks explicitly $R$-symmetry and we 
compute the resulting new contribution to the gaugino mass.
In section \ref{Scalari} we study the contributions to the scalar mass, 
giving a model independent answer, and  
we comment on the corrections to the
messenger masses.
In section \ref{Minimal} we provide two minimal realizations of Yukawa-gauge mediation
and we then conclude in section \ref{conclu}.
We leave to the appendix most of the details of the loop integrals computations.


\section{Setup of Yukawa-gauge mediation}\label{setup}

We would like to study in a model independent way 
the pattern of soft masses generated by a Yukawa-gauge mediated model.
The set up is the following: there is a visible sector which communicates
via gauge interactions to a pair of vector-like and weakly coupled messenger
chiral superfields. The messengers are charged only under the SSM gauge group
(that we take to be  $U(1)$). They interact through a trilinear
superpotential 
with a chiral field, that we call $X$.
This chiral field then couples via superpotential interaction to another chiral operator
$\mathcal{O}$, of dimension two.
The SUSY breaking effects are encoded 
in the one and two point functions of this chiral operator
$\mathcal{O}$.
The resulting visible soft mass terms will be determined as functions of them.

We would like to investigate how these simple assumptions
constrain the allowed visible sector 
spectrum.

The Lagrangian of the model is
\bea
\mathcal{L}&=&\mathcal{L}_{vis}+\int d^4 \theta~ \Phi^{\dagger} e^{V_{SSM}} \Phi+\tilde \Phi^{\dagger} e^{-V_{SSM}} \tilde \Phi
+X^{\dagger} X\nonumber   \\
&&
+
\int d^2 \theta ~m \tilde \Phi \Phi+ \lambda_x X \tilde \Phi \Phi+ \lambda_{o} X \mathcal{O} ~+~ h.c.\,   \label{completeL}
\eea 
Note that it is always possible to  perform a rotation of the phases 
of the different fields and of $\mathcal{O}$ 
in such a way that the mass parameter $m$ and the couplings $\lambda_{x}$ and $\lambda_{o}$ are all real. 
Here we do not add an explicit mass term for $X$ in order to not introduce extra
dimensionful parameters. This is the typical situation in many effective chiral 
models for supersymmetry breaking, where the field $X$ is a pseudo-modulus which acquires 
mass only through loop corrections.

In the limit in which the SUSY breaking hidden sector operator $\mathcal{O}$ decouples from $X$, 
$\lambda_o \to 0$, we are left with a supersymmetric visible sector.
On the other hand, supersymmetry is also restored in the limit $\lambda_{x} \rightarrow 0$, 
where the chiral field $X$ decouples from the messengers. 
Supersymmetry in the SSM can therefore be recovered in two different ways.
This ambiguity is somehow solved by the fact that, 
in the expressions for the visible sector soft terms, the two couplings $\lambda_x$ and $\lambda_o$
always appear in 
pairs.\footnote{Our set up is similar to the semi-direct gauge mediation of \cite{Argurio}.
In semi-direct gauge mediation the messengers are coupled to an hidden gauge field,
which is also coupled to a non supersymmetric current. Here the messengers are coupled to
a chiral field $X$ which is coupled to a non supersymmetric chiral operator $\mathcal{O}$.
The chiral field $X$ plays the role of the hidden gauge field  and 
the chiral operator $\mathcal{O}$  the role of the hidden sector current of semi-direct gauge mediation.}
One could then simply say that visible sector supersymmetry is restored in the limit $\lambda_{x} \lambda_{o} \rightarrow 0 $.
Finally, note that supersymmetry in the SSM is also recovered by sending to infinity the messenger masses.

The supersymmetry breaking effects can be encoded in the one and two point functions of
the chiral operator $\mathcal{O}=(O,\psi^o,F_o)$. Lorentz invariance constrains them to be of the 
form
\be
\begin{aligned}
&\langle O \rangle = O_o \, , \\
& \langle O(p)O(-p) \rangle = G_0(p^2)\, , \\
&\langle \psi^o_{\alpha}(p) \psi^o_{\beta}(-p) \rangle = \epsilon_{\alpha \beta}  G_2(p^2)\, , \\
& \langle O (p) F_o (-p) \rangle =  G_4 (p^2)\, ,  \\
& \langle F_o (p) F_o^{\dagger} (-p)\rangle =-p^2 G_6(p^2) \, ,
\end{aligned}
\qquad
\qquad
\begin{aligned}
&\langle F_o \rangle= f_o \label{onepoint}\, , \\
&\langle O(p)O^{\dagger}(-p) \rangle = G_1(p^2)\, , \\
&\langle \psi^o_{\alpha}(p) \bar \psi^o_{\dot \beta}(-p) \rangle =  p_{\mu} \sigma^{\mu}_{\alpha \dot \beta} G_3(p^2)\, ,\\
&\langle O (p) F_o^{\dagger}(-p) \rangle = G_5(p^2)\, ,\\
 &\langle F_o(p) F_o(-p) \rangle =G_7(p^2) \, ,
 \end{aligned}
\ee
where the unknown functions $G_i(p^2)$ can depend on different scales of the hidden sector.
Here $G_2$, $G_4$ and $G_5$ have mass dimension one, $G_7$ have mass dimension two,
and the others are dimensionless.
If supersymmetry is unbroken 
\be
G_0=G_5=G_7=0 \, , \qquad G_1=G_3=G_6\, , \qquad G_2=G_4 \, .
\ee

Starting from (\ref{completeL}) we can derive the effective Lagrangian for $X$
\bea
 \delta \mathcal{L}_X&=& \lambda_{o} \left( O_o F_x + f_o x + h.c.\right) \nonumber  \\
 && + \lambda_{0}^{2} (  G_6 x^{\dagger}\Box x- i G_3 \bar \psi_x \bar \sigma^{\mu} \pa_{\mu} \psi_x+G_1 F_x F_x^{\dagger}+  \\
&&\qquad +
G_7 x^2+G_0 F_x F_x+  G_5 F_x x^{\dagger}+G_4 x F_x-\frac{1}{2}G_2 \psi_x \psi_x +  h. c. ) \, . \nonumber 
\eea
One point functions for the operator $\mathcal{O}$  typically induce an $F$-term for the field $X$, 
resulting  in minimal gauge mediation.
Hence,
from now on, we assume that the one point functions of 
the operator $\mathcal{O}$ are vanishing,
i.e. $O_o =
f_o=0$.
This can be enforced by a discrete $Z_2$ symmetry of the hidden sector under which
the operator $\mathcal{O}$ is charged. This symmetry is broken by the coupling
of the operator $\mathcal{O}$ to $X$ and to the messengers. However this $Z_2$ is
enough to loop suppress the generation of a tadpole for the $F$-term of $X$.
This discrete symmetry is the analogous of the messenger parity of \cite{GiudiceOld} 
reviewed in GGM \cite{GGM}.

An equivalent  formulation
consists in encoding the SUSY breaking effect 
directly in the two point functions of the chiral field $X$. 
At first order in the insertion of the hidden sector two point functions $G_{i}$, 
the propagators for the field  $X$ are
\be \label{propagators}
\begin{aligned}
&\langle F_x(p)F_{x}(-p)\rangle = \lambda_o^2 G^{*}_{0}(p^{2}) \, ,\\
&\langle \psi_{x\alpha}(p) \psi_{x}^{\beta}(-p)\rangle=  \frac{\lambda_o^2  G^{*}_{2}(p^{2})\delta_{\alpha}^{\beta}}{p^{2}  } \, , \\
&\langle x(p)F_{x}(-p)\rangle = -\frac{\lambda_o^2 G^{*}_{4}(p^{2})}{p^{2} }\, , \\
&\langle x(p)x^{\dagger}(-p)\rangle =-\frac{\lambda_o^2 G_{6}(p^{2})}{p^{2} }\, ,
\end{aligned}
\qquad\qquad 
\begin{aligned}
&\langle F_x(p)F_{x}^{\dagger}(-p)\rangle = \lambda_o^2 G_{1}(p^{2}) \, , \\
&\langle \psi_{x\alpha}(p)\bar \psi_{x\dot\alpha}(-p)\rangle = 
\frac{\lambda_o^2 G_{3}(p^{2})p_{\mu} \sigma^{\mu}_{\alpha\dot\alpha}}{p^{2} }\, ,\\
& \langle x(p)F^{\dagger}_{x}(-p)\rangle = -\frac{\lambda_o^2 G_{5}(p^{2})}{p^{2} }\, , \\
 &\langle x(p)x(-p)\rangle = \frac{\lambda_o^2 G^{*}_{7}(p^{2}) } {p^{4} }\, ,
\end{aligned} 
\ee 
where we have assumed no mass terms for $X$.

Finally, note that in the ansatz (\ref{completeL}) there is 
a $U(1)_R$ symmetry under which the operator $\mathcal{O}$ has $R$-charge two
and the field $X$ has $R$-charge zero. 
This symmetry should be broken down to $Z_2$ in order to
 generate visible majorana gaugino masses.
Here the only sources for 
$R$-symmetry breaking are the  two point functions $G_i$. 
Non-vanishing $G_2, G_4, G_5$ break this $R$-symmetry to $Z_2$, 
$G_0$ breaks it to $Z_4$, while the other $G_i$ preserve it.


\section{Gaugino mass} \label{gaugino}

In this section we compute the two loop contributions to gaugino mass,
showing the presence of gaugino mass screening in
Yukawa-gauge mediation.
Details about computation of the 
loop integrals are reported in the appendix \ref{loop}.
Here we only discuss the different contributions, the non-trivial cancellations that occur  and the resulting soft term. 

\begin{figure}[th]
\begin{center}
\includegraphics[width=\textwidth]{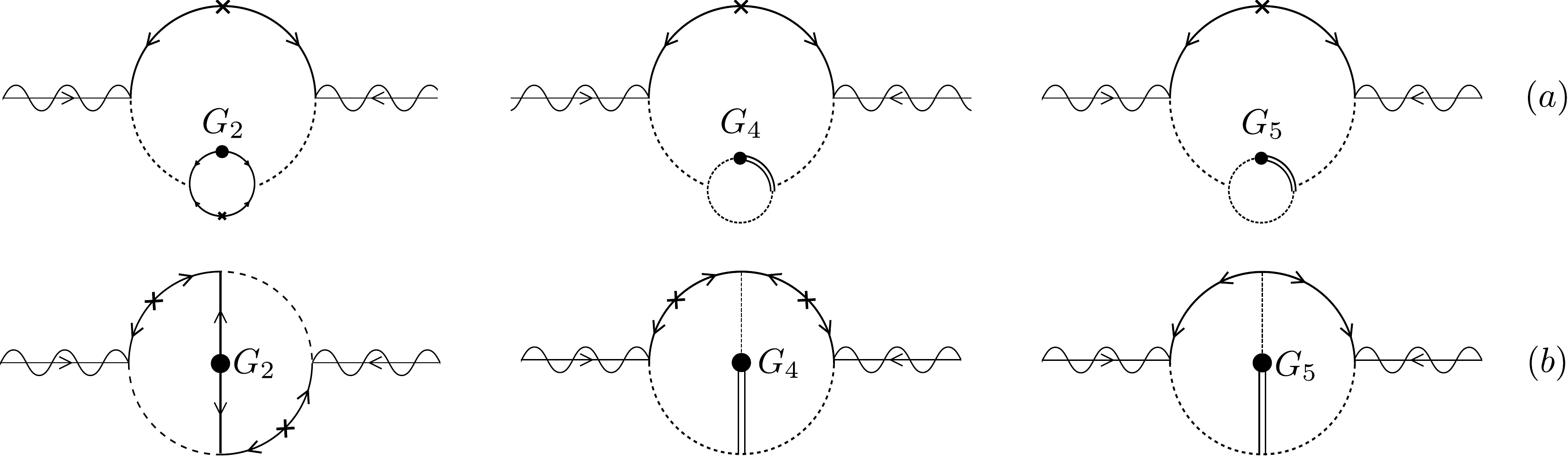}
\end{center}
\caption{The six graphs contributing to visible gaugino mass at leading loop order.  The internal line involving a bubble is a propagator of components of the chiral superfield $X$. All the remaining internal lines correspond to propagators of the messengers.}    
\label{grafici}
\end{figure}

There are six non-equivalent graphs contributing to the gaugino mass.
They are depicted in figure \ref{grafici}.
Note that these involve only the functions $G_2, G_4$ and $G_5$. 
As already observed,
these are indeed the only two point functions 
among \eqref{onepoint} 
which break
 $R$-symmetry down to $Z_2$.

There are two graphs with $G_{5}$
\bea
m_{[a,5]} &=&  8 g^{2} \lambda^{2}_{x} \lambda^{2}_{o}\int \frac{d^{4}k}{(2\pi)^{4}}\int \frac{d^{4}l}{(2\pi)^{4}} \frac{ m^{2}  G_{5}(k^{2})}{k^{2}[l^{2}+m^{2}]^{3}[(l-k)^2+m^2]} \, ,  \\
m_{[b,5]} &=&  -4 g^{2} \lambda^{2}_{x} \lambda^{2}_{o} \int \frac{d^{4}k}{(2\pi)^{4}}\int \frac{d^{4}l}{(2\pi)^{4}} \frac{ l\cdot(l-k) G_{5}(k^{2})}{k^{2}[l^{2}+m^{2}]^{2}[(l-k)^2+m^2]^{2}} \, .
\eea
Performing the analytic integration over the loop momentum $l$ 
it is easy to check that the two contributions cancel each other precisely.\footnote{
The structure of the integrals is exactly as those in \cite{Argurio}.}
We conclude that the two point function $G_5$ does not contribute
to gaugino mass at the leading loop order.

Of the remaining four graphs in figure \ref{grafici}, two involve 
$G_{2}$ and two $G_{4}$. Their explicit expression can be found in appendix \ref{loop}.
They combine in pairs to give the following  integrals in terms of the difference $G_{2}-G_{4}$
\bea
m_{[a]} &=& - 8 g^{2} \lambda^{2}_{x} \lambda^{2}_{o}\int \frac{d^{4}k}{(2\pi)^{4}}\int \frac{d^{4}l}{(2\pi)^{4}} \frac{ m^{2}  \left( G_{2}(k^{2})  -G_{4}(k^{2})\right)}{k^{2}[l^{2}+m^{2}]^{3}[(l-k)^2+m^2]}  \, , \\
m_{[b]} &=&  -4 g^{2} \lambda^{2}_{x} \lambda^{2}_{o} \int \frac{d^{4}k}{(2\pi)^{4}}\int \frac{d^{4}l}{(2\pi)^{4}} \frac{ m^{2}  \left( G_{2}(k^{2})  -G_{4}(k^{2})\right)}{k^{2}[l^{2}+m^{2}]^{2}[(l-k)^2+m^2]^{2}} \, .
\eea
The integration over the loop momentum $l$ can be done analytically and we can write
\be
m_{[a,b]} = - 4 g^{2} \lambda^{2}_{x} \lambda^{2}_{o} \frac{1}{16 \pi^{2}m^{2}} \int \frac{d^{4}k}{(2\pi)^{4}} \frac{1}{k^2}L_{[a,b]}(k^2/m^2)\left( G_{2}(k^{2})  -G_{4}(k^{2})\right)\, ,
\ee
where the kernels $L_{[a,b]}$ are defined and computed in appendix \ref{loop}.
The gaugino mass is then given by
\be \label{gaugmass}
m_{\lambda} = m_{[a]} + m_{[b]} = -\frac{ 4 g^{2} \lambda^{2}_{x} \lambda^{2}_{o}}{16 \pi^{2}} \frac{1}{m^{2}} \int \frac{d^{4}k}{(2\pi)^{4}} \frac{1}{k^2}L(k^2/m^2)\left( G_{2}(k^{2})  -G_{4}(k^{2}) \right)\, ,
\ee
with 
\be \label{gaugmasskernel}
L(x) = L_{[a]}(x)+L_{[b]}(x)=  \frac{x (4+x)+4 \sqrt{x (4+x)} \text{~Arcth}\left(\sqrt{\frac{x}{4+x}}\right)}{x (4+x)^2}\, . 
\ee
This is our general result for the two loop visible gaugino mass
in Yukawa-gauge mediation.

Using a formulation similar to the one in \cite{mugeneral},
we can rewrite this non-vanishing contribution 
as
\be
\label{gauginofin}
m_{\lambda} = - \frac{ g^{2} \lambda^{2}_{x} \lambda^{2}_{o}}{16 \pi^{2}}\frac{1}{m^{2}} \int \frac{d^{4}k}{(2\pi)^{4}} \frac{\bar\sigma^{\mu\dot\alpha\alpha}k_{\mu}}{(k^2)^2}L(k^2/m^2) 
\langle \{ Q_{\alpha},[\bar Q_{\dot \alpha}, F_o O] \} \rangle \, .
\ee
Note that arguments in \cite{mugeneral} show that 
the combination $G_2-G_4 \simeq \langle \{ Q_{\alpha},[\bar Q_{\dot \alpha}, F_o O ] \} \rangle$ 
is subleading in the SUSY breaking scale. 
On the other hand, $G_5$
is at leading order \cite{mugeneral},
but its contribution to the gaugino mass cancels out as we have just discussed.
Hence, whatever the integral over $k^2$, we expect 
the gaugino mass to be suppressed in this class of models.

We conclude that, generically, in Yukawa-gauge mediation the gaugino
mass receives contributions that are at the leading loop order (at least two loops), 
but  subleading in the SUSY breaking scale.
This is in agreement with the gaugino mass screening result derived via
analytic continuation in superspace techniques \cite{Hamed}, 
through which only the leading effects in the SUSY breaking scale are captured.
In our computation we obtain an expression encoding
the next order contributions in the SUSY breaking scale, for the generic
set up explained in section \ref{setup}.


\subsection{Breaking $R$-symmetry explicitly}
The Lagrangian in section \ref{setup} has 
a $U(1)_R$ symmetry under which $R[\mathcal{O}]=2$ and $R[X]=0$.
As a consequence, the two point functions that  can enter into the expression for the visible gaugino mass
are only the ones that break the $R$-symmetry down to $Z_2$. 
Note however that the cancellation of the $G_5$ contribution and the resulting suppression
in the SUSY breaking scale of the gaugino mass is not a direct consequence of $R$-symmetry.

Nevertheless, we could modify the previous scenario by adding an explicit $R$-symmetry
breaking term in the superpotential for the chiral superfield $X$.
The most generic renormalizable superpotential is 
\be \label{def}
\Delta W = \frac{m_{x}}{2} X^{2} + \frac{ \lambda_{def}}{3} X^3\, .
\ee
This breaks completely the $R$-symmetry and, as a consequence, 
other  $G_{i}$  can now enter into the gaugino mass computation.

\begin{figure}[h]
\begin{center}
\includegraphics[width=0.5\textwidth]{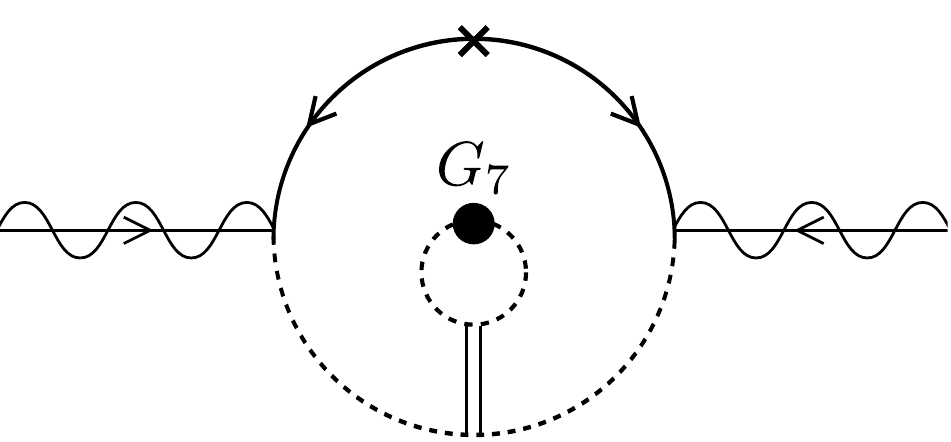}
\end{center}
\caption{The additional graph contributing to visible gaugino mass
with
the 
superpotential deformation $\Delta W$. }
\label{deformati}
\end{figure}

After a careful analysis, one can show that the only extra diagram
that is now generated  at  two loop is the one in figure \ref{deformati}. 
This gives an extra contribution to the gaugino mass 
\bea \label{gauginog7}
\Delta m_{\lambda} &=& 4  g^2 \lambda_{x}\lambda_{def} \lambda^{2}_{o} \int \frac{d^4 k}{(2\pi)^4}\int \frac{d^4 l}{(2\pi)^4} \frac{m G_{7}(k^{2}) }{ [k^2+m_{x}^{2}]^{2}  [l^2+m^2]^{3}} \nonumber \\
&=& \frac{  2 g^2 \lambda_{x}\lambda_{def}\lambda^{2}_{o} }{16 \pi^2}\frac{1}{m} \int \frac{d^4 k}{(2\pi)^4} \frac{G_{7} (k^{2}) }{ [k^2+m_{x}^{2}]^{2} } \, ,
\eea
which is typically not subleading in the SUSY breaking scale.

This is not in disagreement with the result of \cite{Hamed}. In fact, in 
this particular setting,
we are effectively generating a tadpole for the $F$-term of $X$ at one loop
and the contribution
to the gaugino mass is expected to be proportional to this loop generated $F$-term.


\section{Scalar masses} \label{Scalari}
In this section we compute the induced 
soft masses for the scalars of the SSM. 

The model at hand is
a subclass of general gauge mediation \cite{GGM}.  We
can express the scalar masses as a loop integral over a combination of the functions $C_i$
\be \label{GGMscalars}
m_{sf}^2=-g^4 \int \frac{d^4 p}{(2 \pi)^4} \frac{1}{p^2}  (3C_1(p^2)-4 C_{1/2}(p^2)+C_0(p^2)) \, ,
\ee
where the $C_i$ can be computed by considering the 
quantum corrections to the propagators of each component 
of the vector superfield.
Following the  same strategy of  \cite{Argurio}, we can compute the $C_i$ as functions of 
the unknown $G_i$ and, exchanging the order of integration and  performing some of the loop integrals,  
re-express the scalar masses
as  one loop integral of the functions $G_i$  convoluted with a non-trivial kernel.
We argue, following the reasoning in \cite{GMGM},
that the final answer is of the form
\bea
m_{sf}^2 &=&\frac{g^4 \lambda_x^2 \lambda_o^2}{(16 \pi^{2})^2} \int \frac{d^4 k}{(2\pi)^4 k^4} S(k^2/m^2) 
\langle Q^4 \left(  O^{\dagger}  O\right) \rangle= \nonumber \\
\label{massesca}
&=&\frac{g^4 \lambda_x^2 \lambda_o^2}{(16 \pi^{2})^2} \int \frac{d^4 k}{(2\pi)^4 k^2} S(k^2/m^2) (G_1(k^2)-2 G_3(k^2)+G_6(k^2)) \, ,
\eea
thus depending only on the combination $G_1-2 G_3+G_6$.

In order to compute the kernel $S(k^2)$ it is sufficient to focus only on the contribution to  \eqref{GGMscalars}
  given by $G_1$. 
  To every $C_{i}$  corresponds  a two loop integral, involving $G_{1}$, that we denote as  
\be
C^{1}_{i}(p^{2})=   \lambda_x^2 \lambda_o^2 \int  \frac{d^4 l}{(2\pi)^4} \frac{d^4 k}{(2\pi)^4 }  S_{i,1}(l,k,p,m) G_1(k^2) \, .
\ee
These expressions 
and in particular the functions $S_{i,1}$ are computed in appendix \ref{loopscalari}.
Matching the formula \eqref{GGMscalars} and  \eqref{massesca}  
\be
\frac{ \lambda_x^2 \lambda_o^2}{(16 \pi^{2})^2} \int \frac{d^4 k}{(2\pi)^4 k^2} S(k^2) G_1(k^2)=-\int \frac{d^4 p}{(2\pi)^4 p^2} \left(C^{1}_0(p^2)-4 C^{1}_{1/2}(p^2)+3C^{1}_1(p^2)\right) \, ,\nonumber 
\ee
we can identify the kernel 
\bea
\frac{1}{(16 \pi^{2})^2}\frac{S(k^2)}{k^2}= \qquad&& \nonumber  \\
 -\!\!\int \!\!\!\frac{d^4 p}{(2\pi)^4 p^2} \frac{d^4 l}{(2\pi)^4} &\!\!\! &\!\!\!  \large(S_{0,1}(l,k,p,m) -4 S_{1/2,1}(l,k,p,m) + 3 S_{1,1}(l,k,p,m)  \large) \, .
\eea
This integration can be performed only numerically,
 order by order in an expansion in $k^2$.

The final answer is that the kernel $S(k^2/m^2)$ is a positive function of
$k^2/m^2$, with the following 
expansion for small momenta
\be
\label{smallpS}
S(x \to 0)=4x -\frac{4}{9} x^2 +\dots
\ee
Moreover, $S(k^2/m^2)$ behaves logaritmically at large momenta,
ensuring the convergence of the integral.

The expression at small momenta provides a consistency check of our computation.
Indeed, we can recover the case of minimal gauge mediation 
by setting
$G_6=G_3=0$ and $G_1 \simeq |F|^2 (2 \pi)^4\delta^4(k)$. 
In this particular limit the diagrams 
involved in the computation are exactly the diagrams of minimal gauge mediation, at first
order in the supersymmetry breaking parameter $F$. 
Plugging this ansatz in (\ref{massesca})
the integral over $k^2$ can be done trivially and 
our result for $S(k^2/m^2)$ gives 
$m_{sf}^2 \simeq \frac{g^4 \lambda_x^2 \lambda_o^2}{(16 \pi^2)^2} \frac{|F^2|}{|m^2|}$, which coincides with 
minimal gauge mediation
if we consider that the effective $F$-term is $F_{eff}=F \lambda_x \lambda_o$.

We conclude that (\ref{massesca}) is the model independent expression for the
scalar masses in Yukawa-gauge mediation. 
The sign of the scalar masses is then determined 
by the sign of the combination $G_1-2 G_3+G_6$.

\subsection{Messenger mass corrections}\label{StracciaMess}

For completeness we can
compute the corrections to the messenger mass matrix induced
by the coupling with the hidden sector.

\begin{figure}
\begin{center}
\includegraphics[width=0.7\textwidth]{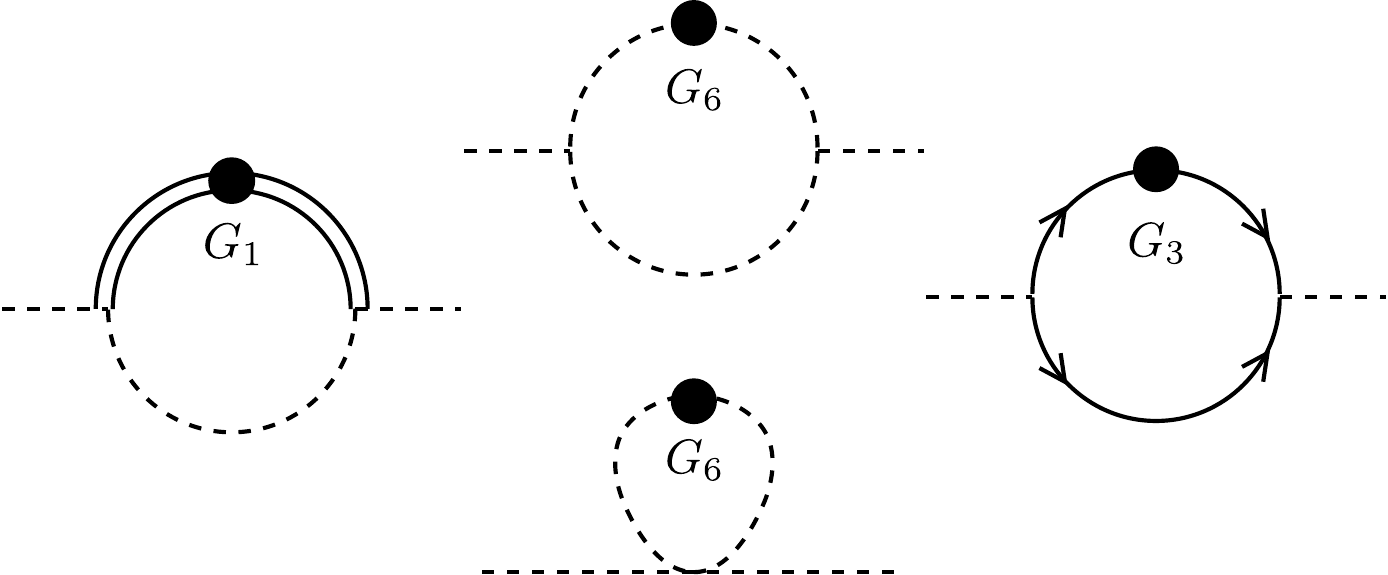}
\end{center}
\caption{ Graphs contributing to the diagonal element of the mass matrix for the messengers.} 
\label{straccia}
\end{figure}

At one loop the diagonal entries of the messenger mass matrix are corrected by the graphs in figure 
\ref{straccia},
which read
\be \label{stracciadiag}
\delta m^2_{\Phi\Phi^{*}} =  -\lambda^{2}_{x} \lambda^{2}_{o} \int \frac{dk^{4}}{(2\pi)^4}\frac{G_{1}-2G_{3}  + G_{6}}{[k^{2}+m^{2}]} \, .
\ee
The supertrace over the messengers sector receives correction proportional to
this contribution.
Notice that it has the opposite sign with respect to the
visible sector scalar squared masses.
This is a common feature of gauge mediated models
with messengers \cite{PoppitzTrivedi}.

The off diagonal one loop correction is 
\be\label{stracciaoutof}
\delta m^2_{\Phi\tilde\Phi} =  2 \lambda^{2}_{x}\lambda^{2}_{o} \int \frac{dk^{4}}{(2\pi)^4}\frac{ m  \left( G_{2} -G_{4} -G_{5}\right) }{k^{2} [k^{2}+m^{2}]  } \, .
\ee
These expressions should be taken into account in explicit models
of Yukawa-gauge mediation to check that the messengers do not
turn tachionic due to quantum corrections.

\section{Examples of Yukawa-gauge mediation} \label{Minimal}

\subsection{Example 1: a toy model}
A very basic model of Yukawa-gauge mediation can be obtained introducing one single extra chiral field $Y$ coupled to a spurion 
$M_0= m_{0}+ \theta^2 f$ and setting $ O =  M Y $. In such a way 
\be
\label{primotoy}
W_{o} = \lambda_{o}M  X Y 
\ee
 has simply the form of a mass term for $X$ and $Y$. The Lagrangian for the chiral field $Y$ is 
\be \label{toy}
\mathcal{L}=\int d^4 \theta ~Y^{\dagger} Y  + \int d^2 \theta~\frac{1}{2} M_0 Y^2  + h.c. \, .
\ee
A parity symmetry for $Y$, that forbids $F$-term generation for  $X$, is present.  
Here we have two supersymmetric scales $M$ and $m_{0}$, together with a SUSY  breaking scale $f$, 
which  we take all real for simplicity. We will consider the SUSY breaking scale $\sqrt{f}$ to be always the smallest.

Note that this is clearly only a toy model since the
interaction of $X$ with the hidden sector is effectively a mass term and, furthermore, 
the hidden sector consists of free chiral fields only.
Nevertheless,  this toy model is sufficient to highlight the basic features of Yukawa-gauge mediation
that we have found with the general analysis,
i.e. a suppression of the two loop gaugino mass in powers of the SUSY breaking scale.

All the functions $G_{i}$ in this model, at the lowest non-vanishing order in the SUSY breaking scale, 
are  generated at tree level. 
The quantities we need to give an estimate of the 
gaugino mass are:
\bea
G_{2}(p^{2})-G_{4}(p^{2}) &= &M^{2}     \frac{m_{0}f^{2} } {[p^{2}+ m_{0}^{2}]^{3}} \, ,  \nonumber \\
G_{5}(p^{2}) &=&- M^{2} \frac{m_{0} f} {[p^{2}+m_{0}^{2}]^{2}} \, , \\
G_{7}(p^{2}) &=& M^{2}   \frac{m_{0}^2 f } {[p^{2}+ m_{0}^{2}]^{2}} \, .  \nonumber 
\eea
We have also written the explicit expression for $G_{5}$ to illustrate that 
it is first order in the SUSY breaking scale. 
However, as proven earlier, this does not  contribute to the visible gaugino mass.

We start from  \eqref{gaugmass}
\be
m_{\lambda} = - 4 g^{2} \lambda^{2}_{x}\lambda^{2}_{o} \frac{1}{16 \pi^{2}m^{2}} \int \frac{d^{4}k}{(2\pi)^{4}} \frac{1}{k^2}L(k^2/m^2)\left( G_{2}(k^{2})  -G_{4}(k^{2}) \right)
\ee
with $L(k^2/m^2)$ given by \eqref{gaugmasskernel}
and we plug in the explicit expression of $G_{2}(p^{2})-G_{4}(p^{2})$.
In the limit  $m \gg m_{0} \gg \sqrt{f}$, the gaugino mass  at leading order evaluates to 
\be \label{gautoy}
m_{\lambda} \simeq  -\frac{g^{2} \lambda^{2}_{x} \lambda^{2}_{o} }{(16\pi^{2})^{2}}\frac{M^{2}}{m^{2}}\frac{f^{2}}{m_{0}^{3}} \, .
\ee
This is at two loop order, but it is at second order in the
SUSY breaking scale $f$. Thus we see the suppression we have explained 
in the previous sections.

In the case $R$-symmetry is  broken explicitly by the deformation \eqref{def}, 
the gaugino mass has the extra correction \eqref{gauginog7} involving $G_{7}$. 
In the same limit as above, with also $m_{x}  \sim m$, this reads
\be
 \Delta m_{\lambda} \simeq -4 \frac{ g^2 \lambda_{x}\lambda_{def} \lambda^{2}_{o}}{(16 \pi^2)^{2}} \frac{M^{2}m^{2}_{0} f}{m m^{4}_{x}} \, . 
\ee
Observe that as expected such a correction is still at two loop order, but 
at leading order in the SUSY breaking scale. 
This dominates over \eqref{gautoy} in the range of parameters $\frac{f}{m^{2}_{0}} < \frac{m^{3}_{0}}{m^{3}}$.

For completeness
we can also give an estimate of the sfermion masses generated
in this toy model.
The combination of $G_i$ functions we need is
\be
\label{perscalari}
G_{1}(p^{2})-2G_{3}(p^{2}) + G_{6}(p^{2}) 
=
\frac{M^2 \left(p^2-m_0^2\right) f^2 }{p^2 \left(m_0^2+p^2\right)^3}
\ee
This expression dies off for momenta larger than $m_0$, that we can
use as an effective cutoff in the convolution with the kernel $S(k^2/m^2)$ in \eqref{massesca}.
In the limit $m\gg m_0$ we can approximate the kernel 
as in \eqref{smallpS} and we find
\be
m_{sf}^2
\sim 
-\frac{g^4 \lambda_x^2 \lambda_o^2}{(16 \pi^2)^3}
\frac{M^2  f^2}{ m^2 m_0^2}
\ee
The sfermion masses are negative in this model, therefore it should be considered only 
as a toy model.
Note however that these masses are generated at second order in $f$.
Hence, besides the loop factor, we find a suppression in powers of the SUSY breaking scale
for the gaugino mass compared with the sfermion masses, as expected.

\subsection{Example 2}
A less basic model of Yukawa-gauge mediation 
is characterized by two extra chiral fields $Y$  and $Z$ coupled to a spurion
$M_0= m_{0}+ \theta^2 f$. 
Precisely we set $O=  Y Z$. 
The superpotential term in \eqref{completeL} is therefore of the form 
\be
W_{o} = \lambda_{o} XYZ\, ,
\ee
without any dimensionful parameter. In this model the coupling of $X$ with the hidden
chiral operator is indeed a Yukawa interaction.

The Lagrangian for the two extra superfields $Y$ and $Z$ is 
\be \label{exe}
\mathcal{L}=\int d^4 \theta~ \left( Y^{\dagger} Y + Z^{\dagger} Z \right) + \int d^2 \theta ~\frac{1}{2} M_0 \left( Y^2  +  Z^{2}\right)+ h.c. \, .
\ee
For simplicity we have introduced  only one supersymmetric scale $m_0$ and the SUSY breaking scale $f$, 
which we take real.
Also this model possesses the discrete parity symmetry discussed in the introduction
that forbids the radiative generation of a tadpole for the $F$-term of $X$.

We now show by computing the gaugino mass at order $\lambda_{o}^2$
that this simple model reproduces the qualitative
features we have highlighted with the general analysis.
At leading order in 
$f$,
 $G_{2}-G_{4}$ 
is given by
\bea
G_{2}(p^{2})-G_{4}(p^{2} )&=& 2  m_{0} \int \frac{dk^4}{(2\pi)^4} \frac{ f^{3} }{[k^2+m_{0}^{2}]^{3}[(k-p)^{2}+m_{0}^{2}]^{2}} \nonumber \\ 
&=& \frac{  f^{3}   }{   16 \pi^2 m_{0}^{5}} g_{2-4}(p^{2}/m_{0}^{2})
\eea
where
\be
 g_{2-4}(x) =  - \frac { x(2-x) (4+x)  - 8  (1+x)\sqrt{x (4+x)} \text{Arcth}\left(\sqrt{\frac{x}{4+x}}\right)}{x^{2} (4+x)^{3}} \, . \\
\ee 
Plugging this expression into  \eqref{gaugmass} with \eqref{gaugmasskernel} one can read off the gaugino mass. 
In the specific case where the messengers mass is larger than 
all the other mass scales in the problem, i.e. $m \gg m_{0} \gg \sqrt{f}$,
\be
\label{gauginomass2}
m_{\lambda} \simeq  - 4    \frac{g^{2} \lambda^{2}_{x} \lambda_{o}^{2}}{(16 \pi^{2})^{3} }  \frac{f^3}{8  m^{2}m_{0}^{3}} \, .
\ee
This contribution is at the leading order in loop factors (three loops),
but suppressed in the SUSY breaking scale up to the third order.
This is even a larger suppression than we would expect on the basis
of the general study in section \ref{gaugino}. It would be nice to
analyze if this is a generic feature of models with only chiral fields and renormalizable interactions 
in the hidden sector.
 
If we introduce also the
deformation $\eqref{def}$, the additional graph involving $G_{7}$ contributes to the gaugino mass. 
At the lowest order in $f$ 
\bea
G_{7}(p^{2})&=& 2 m_{0}   \int \frac{dk^4}{(2\pi)^4} \left( \frac{ f^{2} }{[k^2+m_{0}^{2}]^{2}[(k-p)^{2}+m_{0}^{2}]^{2}} + \frac{ f^{2} }{[k^2+m_{0}^{2}]^{3}[(k-p)^{2}+m_{0}^{2}]} \right) \nonumber \\ 
&=&   \frac{2 f^{2}   }{  16\pi^2 m^{2}_{0}} g_{7}(p^{2}/ m_{0}^{2})\,  , 
\eea
with
\be
g_{7}(x)= \frac{x^2(4+x) +4 x \sqrt{x(4+x)}  \text{~Arcth}\left(\sqrt{\frac{x}{4+x}}\right)}{x^2
   (4+x)^2}\, . 
\ee
We can therefore evaluate the mass of the gaugino 
in this setup with $m_{x} \sim m$
\be
\Delta 
m_{\lambda} \simeq
 \frac{  4 g^2 \lambda_{x} \lambda_{def} \lambda^{2}_{o}  }{(16 \pi^2)^{3}}  \frac{f^{2}   }{    m m^{2}_{x}} \, .
\ee
Note that this contribution is at three loops and at
 second order in the SUSY breaking scale, so it results
suppressed one order less with respect to \eqref{gauginomass2}. 
As in the previous toy model, 
this contribution dominates over the latter
if $\frac{f}{m^{2}_{0}} < \frac{m^{3}_{0}}{m^{3}}$.

\section{Conclusions} \label{conclu}
In this paper we have studied in a model independent formalism a
specific subclass of gauge mediation, which we referred to as Yukawa-gauge mediation.
This is characterized by messenger fields which couple
to a singlet chiral field $X$ which, in turn, couples only via
superpotential interaction to a chiral operator $\mathcal{O}$
that parametrizes the hidden sector.
Assuming a parity symmetry that protects $\mathcal{O}$  from taking vevs,  
the phenomenology is encoded in its two point functions.
This class of models is interesting since they can emerge naturally as
the low energy effective theory of 
dynamical supersymmetry breaking models,
and their phenomenology can be very different from the minimal gauge mediation case.

In this set up we studied the resulting soft masses of the gaugino
and of the scalars of the SSM.
We found that generically the gaugino mass is suppressed in
powers of the SUSY breaking scale.
The scalar masses are instead typically generated at the leading order.
We have explicitly illustrated these features in two explicit examples.

It is interesting to make a parallel between our results on gaugino screening 
and what happens in semi-direct gauge mediation (SDGM) \cite{SemiDirect}.
The two loop diagrammatic cancellation we find for $G_5$ is the same that occurs in SDGM \cite{Argurio}. In SDGM the two loop gaugino mass is vanishing at all orders in the SUSY breaking scale and the next non-vanishing contribution is at higher loop orders.
In Yukawa-gauge mediation, instead, there survives a two loop contribution to the gaugino mass
(proportional to $G_2-G_4$), but this is subleading in an expansion in the SUSY breaking scale.
The gaugino mass  is hence screened in both classes of models, but in a substantially different manner.

In our analysis we also studied  
the case of a superpotential deformation that can
lead to a gaugino mass of the same order in the SUSY breaking scale than the scalar masses.
This realizes a scenario with a hierarchy of one loop factor between gaugino and
sfermion masses.
It would be relevant to investigate if 
this particular feature, which results very difficult to realize in general,
can be obtained in other and more generic setups with
weakly coupled chiral fields.

A strategy that has been proposed in order to avoid gaugino screening in SDGM
is to consider chiral messengers \cite{chiralmess}. 
It would be interesting to reconsider our analysis under this different assumption.
The results of \cite{chiralmess} suggest that 
the gaugino mass would be 
unscreened also in Yukawa-gauge mediation,
but a careful investigation is necessary.

Finally we would like to comment on a different ansatz for Yukawa-gauge mediation.
Instead of parametrizing the coupling of $X$ with the hidden sector as in \eqref{completeL},  one could take a superpotential of the form 
$
W = \lambda_{\tilde o} X X  \mathcal{\tilde O}\, .
$
As long as we consider only one point functions 
of $\mathcal{\tilde O}$,  the general analysis of the previous sections is  valid
and   one can still parametrize the SUSY breaking effects as in \eqref{propagators}.\footnote{The only caveat 
 is in the definition of  the functions $G_{i}$, which must now be computed according to  \eqref{propagators}.}
 Going to the next order in $\lambda_{\tilde o}$ the two point functions of $\mathcal{\tilde O}$ play a role, but we expect that our results similarly extend with minimal adjustments also to this case. 

 More generally, it would be interesting to study the extended setup where $X$ interacts with hidden
sector operators both via linear and quadratic superpotential terms.

\acknowledgments
We are grateful to Riccardo Argurio, Gabriele Ferretti, Dan Thompson and Brian Wecht for
useful comments on the draft.
F.G. is an Aspirant of FWO-Vlaanderen and A.M. is a Postdoctoral Researcher of FWO-Vlaanderen.
This work is supported in part by the FWO-Vlaanderen  through the project G.0114.10N, and in part by the Belgian Federal Science Policy Office through the Interuniversity Attraction Pole IAP VI/11.

\appendix

\section{Gaugino mass loop integrals}\label{loop}

We have six graphs contributing to the gaugino mass.
They are reported in figure \ref{grafici}.
Two involve $G_{2}$. 
The first  of those  gives:
\be 
m_{[a,2]} = - 8 g^{2} \lambda^{2}_{x} \lambda^{2}_{o}  \int \frac{d^{4}k}{(2\pi)^{4}}\int \frac{d^{4}l}{(2\pi)^{4}} \frac{ m^{2}  G_{2}(k^{2})}{k^{2}[l^{2}+m^{2}]^{3}[(l-k)^2+m^2]} \, , \nonumber
\ee  
 The second  contribution with $G_{2}$ is:
\be
m_{[b,2]} = - 4 g^{2} \lambda^{2}_{x} \lambda^{2}_{o}  \int \frac{d^{4}k}{(2\pi)^{4}}\int \frac{d^{4}l}{(2\pi)^{4}} \frac{ m^{2}  G_{2}(k^{2})}{k^{2}[l^{2}+m^{2}]^{2}[(l-k)^2+m^2]^{2}} \, . \nn
\ee 
Other two graphs involve $G_{4}$, they evaluate respectively to
\be
m_{[a,4]} = 8 g^{2} \lambda^{2}_{x} \lambda^{2}_{o}  \int \frac{d^{4}k}{(2\pi)^{4}}\int \frac{d^{4}l}{(2\pi)^{4}} \frac{ m^{2}  G_{4}(k^{2})}{k^{2}[l^{2}+m^{2}]^{3}[(l-k)^2+m^2]} \nn
\ee
and 
\be
m_{[b,4]} = 4 g^{2} \lambda^{2}_{x} \lambda^{2}_{o}  \int \frac{d^{4}k}{(2\pi)^{4}}\int \frac{d^{4}l}{(2\pi)^{4}} \frac{ m^{2} G_{4}(k^{2})}{k^{2}[l^{2}+m^{2}]^{2}[(l-k)^2+m^2]^{2}} \, . \nn
\ee  
The different factor of $2$  in graphs with the $a$ and $b$ topology is due to the internal loops.
In fact, in $m_{[a,2]}$ there is a fermionic loop involving $G_2$, which brings a factor of $2$.
In $m_{[a,4]}$ 
there are two possible equivalent ways to insert the internal loop involving $G_4$ in the scalar propagator.

Finally,  two graphs involving $G_{5}$ contribute to the gaugino mass as
\be
m_{[a,5]} =  8 g^{2} \lambda^{2}_{x} \lambda^{2}_{o}  \int \frac{d^{4}k}{(2\pi)^{4}}\int \frac{d^{4}l}{(2\pi)^{4}} \frac{ m^{2}   G_{5}(k^{2})}{k^{2}[l^{2}+m^{2}]^{3}[(l-k)^2+m^2]} \, \nn
\ee
and
\be
m_{[b,5]} =  -4 g^{2} \lambda^{2}_{x}  \lambda^{2}_{o} \int \frac{d^{4}k}{(2\pi)^{4}}\int \frac{d^{4}l}{(2\pi)^{4}} \frac{ l\cdot(l-k)  G_{5}(k^{2})}{k^{2}[l^{2}+m^{2}]^{2}[(l-k)^2+m^2]^{2}} \, . \nn
\ee  
The mismatching  factor of $2$ is as for $G_{4}$. 
The difference in sign
for $m_{[b,5]}$ 
is because of the fermion propagators, 
that involve sigma-matrices and not masses, as for $m_{[b,4]}$. 
As explained in the main text,  these two contributions cancel each other as in \cite{Argurio}, 
giving no correction to the visible gaugino mass.

There are therefore four graphs that contributes to the gaugino mass. They combine in pairs to give the following  integrals in terms of the difference $G_{2}-G_{4}$
\be
m_{[a]} = - 8 g^{2} \lambda^{2}_{x} \lambda^{2}_{o}  \int \frac{d^{4}k}{(2\pi)^{4}}\int \frac{d^{4}l}{(2\pi)^{4}} \frac{ m^{2}  \left( G_{2}(k^{2})  -G_{4}(k^{2}) \right)}{k^{2}[l^{2}+m^{2}]^{3}[(l-k)^2+m^2]}  \, , \nn 
\ee
\be
m_{[b]} =  -4 g^{2} \lambda^{2}_{x} \lambda^{2}_{o}  \int \frac{d^{4}k}{(2\pi)^{4}}\int \frac{d^{4}l}{(2\pi)^{4}} \frac{ m^{2}  \left( G_{2}(k^{2})  -G_{4}(k^{2})\right)}{k^{2}[l^{2}+m^{2}]^{2}[(l-k)^2+m^2]^{2}} \, . \nn
\ee 
The integration over the $l$ momentum can be done analytically and we can write
\be
m_{[a,b]} = - 4 g^{2} \lambda^{2}_{x}  \lambda^{2}_{o} \frac{1}{ 16 \pi^{2} m^{2}} \int \frac{d^{4}k}{(2\pi)^{4}} \frac{1}{k^2}L_{[a,b]}(k^2/m^2)\left( G_{2}(k^{2})  -G_{4}(k^{2})\right)\, , \nn
\ee
where
\bea
L_{[a]}(k^2/m^2) &=& 2 \int  d^{4}l ~ \frac{ m^{4} }{[l^{2}+m^{2}]^{3}[(l-k)^2+m^2]} \, , \nn\\
L_{[b]}(k^2/m^2) &=& - \int d^{4}l  ~\frac{  m^{4} }{[l^{2}+m^{2}]^{2}[(l-k)^2+m^2]^{2}} \, . \nn
\eea
Using the Feynman parametrization,  exchanging the order of integration between the loop integral and the Feynman parameter integral  one finally obtains 
\bea
L_{[a]}(x) &=&  \frac{ x(2+x)(4+x)   -  8 \sqrt{x(4+x)}  \text{~Arcth}  \left(\sqrt{ \frac{x}{4+x} } \right) }{ x^2(4+x)^2}\, , \nn \\
L_{[b]}(x) &=& \frac{ 2x (4+x)-4 (2+x) \sqrt{x (4+x)}  \text{~Arcth} \left(\sqrt{\frac{x}{4+x}}\right)}{x^2 (4+x)^2} \, .  \nn
\eea
The gaugino mass is then given by
\be
m_{\lambda} =  - 4 g^{2} \lambda^{2}_{x} \lambda^{2}_{o} \frac{1}{16\pi^2}  \frac{M}{m^{2}} \int \frac{d^{4}k}{(2\pi)^{4}} \frac{1}{k^2}L(k^2/m^2)\left( G_{2}(k^{2}/M^{2})  -G_{4}(k^{2}/M^{2}) \right) \, , \nn
\ee
with 
\be
L(x) = L_{[a]}+L_{[b]}= \frac{x (4+x)-4 \sqrt{x (4+x)} \text{~Arcth}\left(\sqrt{\frac{x}{4+x}}\right)}{x^2 (4+x)} \, . \nn
\ee

 \section{Computation of the kernel $S(k^2/m^2)$} \label{loopscalari}

 We consider the contribution to the $C_i$ functions , i.e. the propagator of
 the component of the vector superfields, given by the $G_1$ function.
\begin{figure} [th]
\begin{center}
\includegraphics[width=0.8\textwidth]{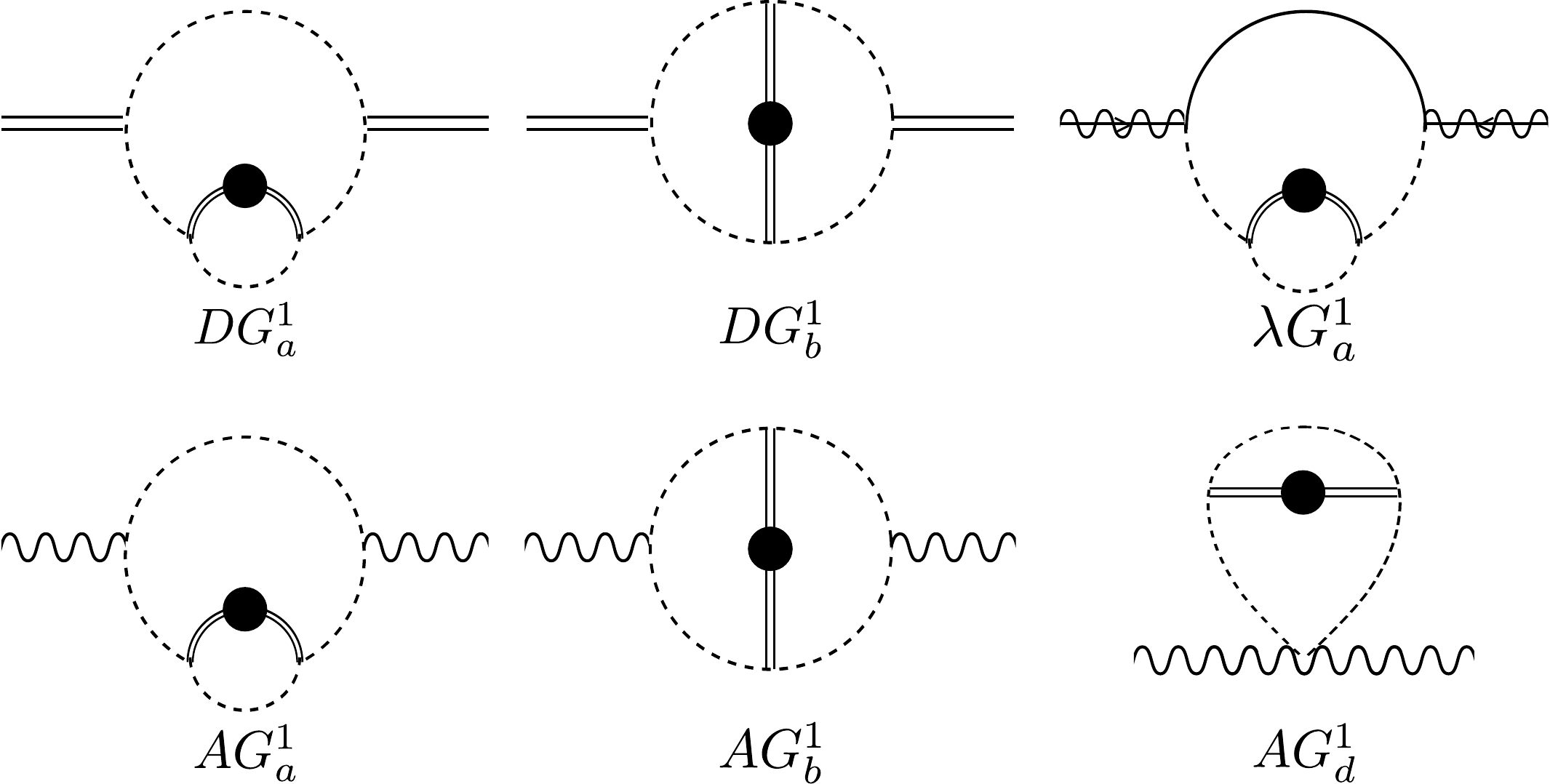}
\end{center}
\caption{Graphs in which $G_{1}$ is contributiong to the visible sfermion masses.}    
\label{fig:scalari}
\end{figure}
 The Feynman diagrams are the one in figure \ref{fig:scalari}, and they evaluate to 

 \bea
\label{diagk0}
DG^1_a &=&4~ \int  \frac{dk^4}{(2\pi)^4} \int \frac{dl^4}{(2\pi)^4} ~\frac{G_1(k^2)}{(l^2+m^2)^2 \, [(l-k)^2 + m^2]\,[(l-p)^2 + m^2]} \, , \nn \\
DG^1_b &=&-2~ \int  \frac{dk^4}{(2\pi)^4} \int \frac{dl^4}{(2\pi)^4} ~\frac{G_1(k^2)}{(l^2+m^2)\,[(l-k)^2 + m^2]\,[(l-k-p)^2 + m^2]\,[(l-p)^2 + m^2]}  \, ,
\nonumber \\  
\lambda G^1_a &=& - 4~\int  \frac{dk^4}{(2\pi)^4} \int \frac{dl^4}{(2\pi)^4} ~ \frac{(l-p)_\mu \,\sigma^\mu_{\alpha \dot\alpha} ~ G_1(k^2)}
{(l^2+m^2)^2 \, [(l-k)^2 + m^2]\,[(l-p)^2 + m^2]}\, ,
 \nn \\
A G^1_a &=& 4~\int  \frac{dk^4}{(2\pi)^4} \int \frac{dl^4}{(2\pi)^4} ~\frac{(2l-p) _\mu (2l-p)_\nu ~ G_1(k^2)}{(l^2+m^2)^2 [(l-k)^2 + m^2]\,[(l-p)^2 + m^2]}\, ,
\\
A G^1_b &=& 2~\int  \frac{dk^4}{(2\pi)^4} \int \frac{dl^4}{(2\pi)^4} ~\frac{(2l-p) _\mu (2l-2k-p)_\nu ~ G_1(k^2)}
{(l^2+m^2)~ [(l-k)^2 + m^2]\,[(l-k-p)^2 + m^2]\,[(l-p)^2 + m^2]}\, , \nn \\
A G^1_d &=& - 4~\int  \frac{dk^4}{(2\pi)^4} \int \frac{dl^4}{(2\pi)^4} ~\frac{\eta_{\mu\nu} ~ G_1(k^2)}{(l^2+m^2)^2 \,[(l-k)^2 + m^2]} \, .\nn
\eea
The functions $C_{i}^{1}$  are simple combinations of these expressions, for instance 
$C^{1}_{0} = \lambda_{o}^{2} \lambda^{2}_{x} \left( DG^1_a+DG^1_{b} \right)$. From this we can obtain the functions $S_{i,1}$ and
the kernel $S(k^2/m^2)$. The terms in \eqref{diagk0} are exactly as in \cite{Argurio}, 
but with a sign difference in $DG^1_b$, and the computation of the kernel can be done along the same
lines.

\end{document}